\documentclass{elsart}
\journal{ \bf This is a draft version dated\ }

\renewcommand{\theequation}{\thesection.\arabic{equation}}
\usepackage[centertags]{amsmath}
\usepackage{newlfont}

\usepackage{graphicx}
\usepackage{psfrag}

\begin{document}
\begin{frontmatter}
\title{Formulas for the amplitude \\ of the van der Pol limit cycle}

\author[Pamp]{J.L. L\'{o}pez},
\author[Tehran,Ghazvin]{S. Abbasbandy},
\author[Zara]{R. L\'{o}pez-Ruiz}

\address[Pamp]{Department of Mathematical and Informatics Engineering,
  Universidad P\'ublica de Navarra, 31006-Pamplona, Spain}

\address[Tehran]{Department of Mathematics, Science and Research
 Branch, Islamic Azad University, Tehran, 14778, Iran}

\address[Ghazvin]{Department of Mathematics, Imam Khomeini
International University, Ghazvin, 34149-16818, Iran}

\address[Zara]{Department of Computer Science and BIFI,
  Universidad de Zaragoza, 50009-Zaragoza, Spain}

\date{\today}

\begin{abstract}
The limit cycle of the van der Pol oscillator,
$\ddot{x}+ \epsilon (x^2-1) \dot{x} + x =0$, is studied in the plane $(x,\dot{x})$
by applying the homotopy analysis method.
A recursive set of formulas that approximate the amplitude and form of this limit cycle
for the whole range of the parameter $\epsilon$ is obtained.
These formulas generate the amplitude with an error less than $0.1\%$.
To our knowledge, this is the first time where an analytical approximation
of the amplitude of the van der Pol limit cycle, with validity from
the weakly up to the strongly nonlinear regime, is given.
\end{abstract}

\end{frontmatter}

\section{Introduction}
\setcounter{equation}{0}

The differential equation,
\begin{equation}
\ddot{x}(t) + \epsilon (x^2-1) \dot{x}(t) + x(t) =0,\ \ \ t \geq 0,
\label{vanderpol}
\end{equation}
with $\epsilon$ a real parameter and the dot denoting the derivative with respect to time $t$,
is called the van der Pol oscillator \cite{vanderpol}.
For $\epsilon>0$, and due to the nonlinear term $\epsilon (x^2-1)\dot{x}$,
the system accumulates energy in the region $\mid x\mid<1$ and dissipates this energy
in the region $\mid x\mid>1$. This constraint implies the existence of an stable
periodic motion ({\it limit cycle} \cite{andronov}) when $\epsilon>0$.
If the nonlinearity is increased, the dynamics in the time domain
runs from near-harmonic oscillations when $\epsilon\rightarrow 0$
to relaxation oscillations when $\epsilon\rightarrow\infty$, making
it a good model for many practical situations \cite{vanderpol1,lopezruiz}.
The closed curve representing this oscillation in the plane $(x,\dot{x})$
is quasi circular when $\epsilon\rightarrow 0$
and a sharp figure when $\epsilon\rightarrow\infty$.
For $\epsilon<0$, the dynamics is dissipative in the region $\mid x\mid<1$ and
amplificative for $\mid x\mid>1$. Under these conditions,
the periodic motion is still possible but unstable. In this case, the limit cycle
can be derived from that one with $\epsilon>0$ taking into account the symmetry
$(\epsilon,x(t))\rightarrow (-\epsilon,-x(-t))$. Therefore, it is enough
to study the case $\epsilon>0$ to obtain also the behavior of the system for $\epsilon<0$.

Different standard methods (perturbative, non-perturbative,
geometrical)
\cite{vanderpol,andronov,vanderpol1,lopezruiz,odani,jordan,ye}
have permitted to study extensively the limit cycle of the van der
Pol equation, in the weakly ($\epsilon\rightarrow 0$) and in the
strongly ($\epsilon\rightarrow\infty$) nonlinear regimes. However,
investigations giving analytical information of this object in the
intermediate  regime ($0\ll \epsilon\ll \infty$) are lacking in
the literature. In this paper, it is our aim to fill in this gap
by applying to the Eq. (\ref{vanderpol}) the homotopy analysis
method (HAM) introduced by Liao \cite{li1,li2} in the nineties.
This method has been shown to be powerful to solve different
nonlinear problems \cite{li3,wu,hayat,ab1,ab2,ab3,tan}. In
particular, it has been applied in \cite{ab4} to Li\'enard
equation, $\ddot{x}+ \epsilon f(x) \dot{x} + x =0$, which is the
generalization of the van der Pol system when $f(x)$ is an
arbitrary function. As the interest in that work \cite{ab4} was
the amplitude and the frequency of the periodic motions, the
calculations were performed with the time variable being explicit.
As here we are only interested in the amplitude and form of the
limit cycles, the time dependence of the solutions can be omitted,
and we can work directly in the phase space $(x,\dot{x})$. Our
results for the van der Pol limit cycle are presented in Section
2. In Section 3 it is explained how these results have been
obtained from our specific application of the HAM to the van der Pol
equation. Some conclusions are given in the last Section.

\section{The amplitude of the van der Pol limit cycle}
\setcounter{equation}{0}

Nowadays, the amplitude of the van der Pol limit cycle $a$ can be easily
approximated with a classical Runge-Kutta method. The results of this computational
calculation $a_E$ are shown in Table 1.
Let us observe that $a_E(\epsilon)>2$ for $\epsilon>0$, with the asymptotic values:
$a_E(\epsilon\rightarrow 0)=a_E(\epsilon\rightarrow \infty)=2$.
An upper bound rigorously established in \cite{odani1} for the amplitude $a$ is $2.3233$.
However, as it was signalled there \cite{odani1}, the maximum of $a_E$ is $2.02342$
and it is obtained for $\epsilon=3.3$. In view of this result, Odani \cite{odani1}
conjectured that {\it the amplitude of the limit cycle of the van der Pol equation is estimated
by $2<a(\epsilon)<2.0235$ for every $\epsilon>0$}.

\begin{center}
\begin{normalsize}
\begin{flushleft}
\begin{tabular}[t]{|p{.6cm}|p{2cm}|p{2cm}|p{2cm}|p{2cm}|p{2cm}|l|l|}
\hline $\epsilon$ & $0.1$ & $0.5$ & $1.0$ & $1.5$ & $2.0$ & $2.5$ \\
\hline $a_{E}$&2.00010&2.00249&2.00862&2.01522&2.01989&2.02235\\
\hline
\hline $\epsilon$ & $3.0$ & $3.5$ & $4.0$ & $5.0$ & $10$ & $50$\\
\hline $a_{E}$&2.02330&2.02337&2.02296&2.02151&2.01428&2.00295\\
\hline
\end{tabular}

\noindent{\bf Table 1.} The value $a_{E}$ represents the amplitude $a$
of the van der Pol limit cycle obtained by integrating directly Eq. (\ref{vanderpol})
with a Runge-Kutta method  for the indicated values of $\epsilon$.
\label{table1}
\end{flushleft}
\end{normalsize}
\end{center}

Due to the difficulty of calculating the amplitude $a$ by analytical means,
one could propose as a good solution the constant amplitude
\begin{equation}
\bar{a}=2
\end{equation}
for the whole range of the parameter $\epsilon$.
Knowing that the experimental upper bound for $a$ is 2.02342, i.e. $2<a(\epsilon)<2.02342$
for every $\epsilon>0$, the error made with this approximation, $(a(\epsilon)-\bar{a})/a(\epsilon)$,
would be about $1\%$.

If more precision is needed, the different analytical expansions in $\epsilon$
that have been found for the amplitude in the weakly and in the strongly nonlinear
regimes can be considered. Evidently, the error becomes very large in the regions
where these approximations are not valid.
In \cite{ab4,ll1,ll2}, a recursive perturbation approximation is used to find the formula
for the amplitude when $\epsilon\rightarrow 0$. This is, up to order $\cal{O}(\epsilon^8)$,
\begin{equation} \label{lla}
a(\epsilon)_{\epsilon\rightarrow 0} = 2+{1\over 96}\ \epsilon^2-{1033\over 552960}\
\epsilon^4+ {1019689\over 55738368000}\ \epsilon^6+\cal{O}(\epsilon^8).
\end{equation}
This expansion agrees for small $\epsilon$ with
the computational calculation $a_E$ presented in Table 1.
For $0<\epsilon<2$ the error is less than $1\%$, for $\epsilon\approx 4$ the error is bigger
than $10\%$, and for $\epsilon\approx 6$ the formula has lost its validity and the error is
bigger than $50\%$.
In \cite{cartwright}, the asymptotic dependence of the amplitude on $\epsilon$
is given for sufficiently large $\epsilon>0$,
\begin{equation} \label{lla1}
a(\epsilon)_{\epsilon\rightarrow \infty}=2+0.7793\ \epsilon^{-4/3}+\underline{\cal{O}}(\epsilon^{-4/3}),
\end{equation}
where $\underline{\cal{O}}$ means $\underline{\cal{O}}(\epsilon^{-4/3})<\cal{O}(\epsilon^{-4/3})$.
Compared with $a_E$, this formula generates the amplitude with an error bigger
than $15\%$ for $\epsilon<2$. The formula starts to have validity for $\epsilon\approx 10$ with an error
around $1\%$, that passes to be less than $0.1\%$ when $\epsilon>50$.
In Figs. 1.a and 1.b, formulas (\ref{lla}) and (\ref{lla1}) are respectively plotted versus $\epsilon$
in their regions of validity.

\vskip 0.5cm
\begin{center}
\begin{figure}[t]
\centerline{\includegraphics[width=7cm]{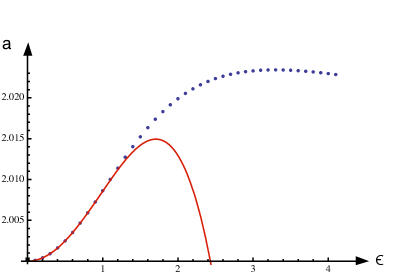}\hskip 5mm\includegraphics[width=7cm]{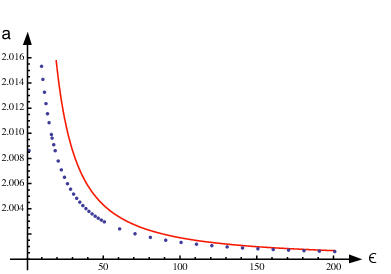}}
\centerline{(a)\hskip 7cm (b)}
\caption{ Comparison of the `experimental' amplitude $a_E(\epsilon)$ (dotted curve) with
{\bf(a)} amplitude $a(\epsilon)_{\epsilon\rightarrow 0}$ given by formula (\ref{lla}) (solid line),
and  {\bf(b)} amplitude $a(\epsilon)_{\epsilon\rightarrow \infty}$ given by formula
(\ref{lla1}) (solid line).}
\end{figure}
\end{center}

We propose here two formulas, $a_{R1}(\epsilon)$ and $a_{R2}(\epsilon)$,
that approximate the amplitude of the van der Pol limit cycle
for every $\epsilon>0$. The first one with an error less than $0.1\%$,
and the second one with an error less than $0.05\%$. These formulas have been obtained
by applying the HAM to the van der Pol equation. The details of the derivation of these formulas
are given in Section 3.

The first formula is
\begin{equation} \label{ar1}
a_{R1}(\epsilon)=
2 + {1.737\ \epsilon^2\over(8 \pi + 9\epsilon) (4 + \epsilon^2)}\ ,
\end{equation}
that derives from expression (\ref{vpauno}). The error obtained,
$|a(\epsilon)-a_{R1}(\epsilon)|/a(\epsilon)$, with this formula is less than $0.1\%$
for every $\epsilon>0$. Fig. 2.1 shows $a_{R1}$ in comparison with the `experimental'
amplitude $a_E$. The maximum of $a_{R1}$ is $2.02317$ and it is taken for $\epsilon=3.3$.

The second formula is
\begin{equation} \label{ar2}
a_{R2}(\epsilon) =
2 + {0.74958\ \epsilon^2\over(8 \pi + 9\epsilon) (4 + \epsilon^2)}\ +
{\epsilon^2(75.3562+43.0023\epsilon+28.1589\epsilon^2+8.3479\epsilon^3)\over
(8\pi + 9\epsilon)^2 (4 + \epsilon^2)^2}\ ,
\end{equation}
that derives from expression (\ref{vpados}). The error obtained,
$|a(\epsilon)-a_{R2}(\epsilon)|/a(\epsilon)$, with this formula is less than $0.05\%$
for every $\epsilon>0$. The maximum of $a_{R2}$ is $2.02346$ and it is taken for $\epsilon=3.29482$.
The plot of $a_{R2}$ in comparison with the amplitude $a_E$ obtained computationally
can be seen in Fig. 2b. Let us observe that it can be guessed from Fig. 2b
that $a_{R2}(\epsilon)>a_E(\epsilon)$ for every $\epsilon>0$.

The expansion of expression (\ref{ar2}) for small $\epsilon$ gives
\begin{equation} \label{ar21}
a_{R2}(\epsilon)_{\epsilon\rightarrow 0}=2 + 0.01491\ \epsilon^2 + {\mathcal O}(\epsilon^3).
\end{equation}
For large $\epsilon$, we obtain
\begin{equation} \label{ar22}
a_{R2}(\epsilon)_{\epsilon\rightarrow \infty}=2 + {0.18648\ \epsilon^{-1}}+{\mathcal O}(\epsilon^{-2}).
\end{equation}
Let us note the different scaling of this last expression (with behavior $\epsilon^{-1}$) respect
to expansion (\ref{lla1}) (with behavior $\epsilon^{-1.33}$).
Taking into account that these approximations start to be valid when $\epsilon>100$, it can be
easily seen that this difference is negligible, in fact less than $0.01\%$.
Nevertheless, we have tried to modify formula (\ref{ar2}) to obtain the correct scaling $\epsilon^{-1.33}$
of expression (\ref{lla1}), but the increase of the error in other regions of the parameter $\epsilon$
does not recommend this possibility.

In summary, let us remark the exceptional fit of the `experimental' points $a_E$ by the amplitudes
$a_{R1}$ and $a_{R2}$ generated with formulas (\ref{ar1}) and (\ref{ar2}),
respectively (see Figs. 2a and 2b). Moreover, by inspection of Fig. 2b, let us finish this section
by posing the following

{\bf Conjecture}: {\it the amplitude $a_{R2}(\epsilon)$ obtained with formula (\ref{ar2}) is an
upper bound for the amplitude $a(\epsilon)$ of the van der Pol limit cycle, that is
\begin{equation} \label{conjec}
a_{R2}(\epsilon)>a(\epsilon) \ \ for\ every\ \ \epsilon>0.
\end{equation}
}

\vskip 0.5cm
\begin{center}
\begin{figure}[t]
\centerline{\includegraphics[width=7cm]{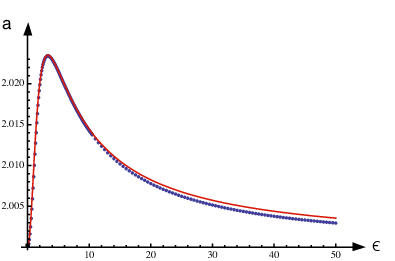}\hskip 5mm\includegraphics[width=7cm]{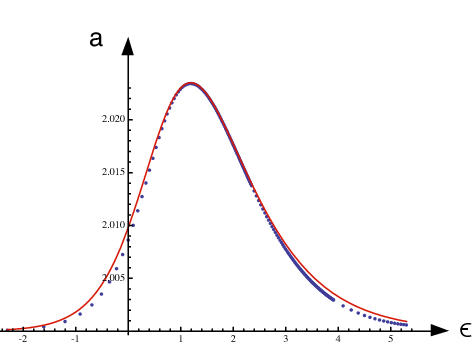}}
\centerline{(a)\hskip 7cm (b)}
\caption{{\bf(a)} Amplitude $a_{R1}(\epsilon)$ given by formula (\ref{ar1}) (solid curve),
and experimental amplitude $a_E(\epsilon)$.
{\bf(b)} Amplitude $a_{R2}(\epsilon)$ given by formula (\ref{ar2}) (solid curve),
and experimental amplitude $a_E(\epsilon)$ (dotted curve), and $\epsilon$ given in logarithmic scale.}
\end{figure}
\end{center}

\section{The HAM and the van der Pol equation}
\setcounter{equation}{0}

The generalization of the van der Pol equation,
$\ddot{u}+ \epsilon (u^2-1) \dot{u} + u =0$,
is called the Li\'enard equation. In coordinates $(u,v)$, it reads
\begin{equation}
\ddot{u}(t) + \epsilon f(u) \dot{u}(t) + u(t) =0,\ \ \ t \geq 0,
\label{lienard}
\end{equation}
where we consider that $f(u)$ is an even function.
The HAM has been applied to this equation in \cite{ab4} working explicitly
in the time domain. We proceed now to apply the HAM to this system
by following a different strategy, namely, by omitting the time variable
of Eq. (\ref{lienard}).

\subsection{HAM applied to Li\'enard equation}

If we define the variable $v=\dot{u}$, and suppose that $v$ is a function of $u$,
the acceleration $\ddot{u}$ can be rewritten as $v'v$, where the prime denotes the
derivative respect to $u$.  The result is a new equation with the time variable eliminated:
\begin{equation}
vv'+\epsilon f(u) v + u=0.
\end{equation}
When $f(u)$ is even \cite{ll3,ll4}, the limit cycles of this equation are closed trajectories
of amplitude $a$ in the $(u,v)$ plane, with $-a\le u\le a$ and $v(a)=v(-a)=0$.

After the re-scaling $(u,v)=(ax,ay)$ to the new variables $(x,y)$, the limit cycles are
transformed in solutions of the equation
\begin{equation}
\label{liena}
yy'+\epsilon y f(ax)+x=0, \hskip 1cm -1\le x\le 1, \hskip 1cm y(-1)=y(1)=0,
\end{equation}
where the prime is now denoting the derivative respect to $x$.

From \cite{ll3,ll4}, we know that an approximate solution of the positive branch solution of ($\ref{liena}$),
for any value of $\epsilon>0$, is
\begin{equation}
\tilde{y}(x)=\sqrt{1-x^2}+{\epsilon\over \tilde a}\left\lbrace
\begin{array}{ll}
0 \hskip 2cm & {if}\hskip 2mm -1\le x<x_0/\tilde a \\
F(\tilde a)-F(\tilde a x) \hskip 5mm & {if}\hskip 2mm x_0/\tilde a\le x\le 1,
\end{array}
\right.
\end{equation}
where $F'(x)=f(x)$, $x_0$ is a negative root of $f(x)$: $f(x_0)=0$ with $x_0<0$, and $\tilde a$ is
a positive root of $F(x_0)-F(x)$: $F(x_0)=F(\tilde a)$ with $\tilde a>0$.
The number of possible stable limit cycles in the strongly non-linear regime (for large $\epsilon$)
is obtained from the number of positive roots of $F(x_0)-F(x)$ when $x_0$ runs over the set of all
the negative roots of $f(x_0)=0$. We take the positive roots $\tilde a$ selected by the algorithm explained
in \cite{ll3,ll4} as the corresponding approximate amplitudes. Let us observe that the initial approximate
limit cycle $\tilde{y}(x)$ recovers the exact circular form when $\epsilon\rightarrow 0$, and also,
when $\epsilon\rightarrow\infty$, it becomes exactly the two-piecewise limit cycle proposed in \cite{ll3}.
Moreover, it must be noticed that $\tilde{y}(x)$ has sense only when $\epsilon>0$, and then the results
obtained from this initial guess only will have validity for $\epsilon>0$.

After eliminating the time variable, the non-linear differential equation $yy'+\epsilon y f(ax)+x=0$
is of order one. Then, to construct the homotopy \cite{li1,li2},
we build a linear differential equation of order one for the whole range $[0,1]$ of what is called
the homotopy parameter $p$.

We define an auxiliary linear operator ${\cal L}$ by
\begin{equation} \label{eq:24}
{\cal L} [\phi(x;p)] =  \frac {\partial}{\partial x} \phi(x;p),
\end{equation}
with the property
\begin{equation} \label{eq:25}
{\cal L} [C_1 ]=0,
\end{equation}
where $C_1$ is a constant, and $p$ is a (homotopy) parameter
explained below.

From (\ref{liena}) we define a nonlinear operator
\begin{equation} \label{nlo}
 {\cal N}[\phi(x;p),A(p)]=
\phi(x;p) \frac{\partial \phi(x;p)}{\partial x} + \epsilon
\phi(x;p) f(A(p)  x) + x,
\end{equation}
and then construct the homotopy
\begin{equation} \label{homo}
{\cal H}[\phi(x;p),A(p)]=(1-p){\cal L} [\phi(x;p)-\tilde{y}(x)] +h
p {\cal N}[\phi(x;p),A(p)],
\end{equation}
where $h$ is a nonzero auxiliary parameter. Setting ${\cal
H}[\phi(x;p),A(p)]=0$, we have the zero-order deformation equation
\begin{equation} \label{zod}
(1-p){\cal L} [\phi(x;p)-\tilde{y}(x)]+h p {\cal N}
[\phi(x;p),A(p)]=0,
\end{equation}
subject to the boundary conditions
\begin{equation} \label{phiz}
\phi(-1;p)=0,\qquad \phi(1;p)=0,
\end{equation}
where $p\in [0,1]$ is an embedding parameter. When the parameter
$p$ increases from 0 to 1, the solution $\phi(x;p)$ varies from
$y_0(x)=\tilde{y}(x)$ to $y(x)$, $A(p)$ varies from $a_0$ to $a$. Assume
that $\phi(x;p)$ and $A(p)$ are analytic in $p\in[0,1]$ and can be
expanded in the Maclaurin series of $p$ as follows:
\begin{equation} \label{mac}
\phi(x;p)=\sum_{n=0}^{+\infty} y_n(x)p^n,\ \
A(p)=\sum_{n=0}^{+\infty} a_n p^n,
\end{equation}
where
\begin{equation*}
y_n(x)=\frac{1}{n!} \frac{\partial ^n \phi(x;p)}{\partial
p^n}\Big|_{p=0},\qquad a_n=\frac{1}{n!} \frac{\partial^n
A(p)}{\partial p^n}\Big|_{p=0}.
\end{equation*}
Notice that series (\ref{mac}) contain the auxiliary parameter
$h$, which has influence on their convergence regions. Assume that
$h$ is properly chosen such that all of these Maclaurin series are
convergent at $p=1$. Hence at $p=1$ we have
\begin{equation*}
y(x)=y_0(x)+\sum_{n=1}^{+\infty} y_n(x),\qquad
a=a_0+\sum_{n=1}^{+\infty} a_n.
\end{equation*}
At the $N$th-order approximation, we have the analytic solution of
Eq.~(\ref{liena}), namely
\begin{equation} \label{eq:14}
y(x) \approx Y_N(x):=\sum_{n=0}^N y_n(x),\ \ a \approx A_N :=
\sum_{n=0}^N a_n.
\end{equation}

The parameter $h$ is free and can be chosen arbitrarily, in
particular, it can be a function of $\epsilon$. Nevertheless, the
function $\tilde y(x)$ is the exact limit cycle solution of the
Li\'enard equation (\ref{liena}) in the limits $\epsilon\to 0$ and
$\epsilon\to\infty$. This means that the general solution $y(x,h)$
of Eq. (\ref{zod}) should tend to $\tilde{y}(x)$ for any $p$ in
those limits of $\epsilon$, loosing its dependence on $h$. Then, a
reasonable property for $h$ would be to vanish in those limits of
$\epsilon$. Hence, the solution of (\ref{zod}) would be the exact
solution $\tilde y(x)$ for any value of the parameter $p$ in the
limits  $\epsilon\to 0$ and $\epsilon\to\infty$. A simple function
$h(\epsilon)$ satisfying these properties is
\begin{equation}
\label{ache}
h(\epsilon)={b\epsilon\over c+\epsilon^2}, \hskip 0.5cm with \hskip 0.5cm b,c\in R.
\end{equation}

Differentiating Eqs.~(\ref{zod}) and (\ref{phiz}) $n$ times with
respect to $p$, then setting $p=0$, and finally dividing by
$n!\,$, we obtain the $n$th-order deformation equation
\begin{equation} \label{mod}
{\cal L} [y_n(x)-\chi_n y_{n-1}(x)]+h R_n(x)=0, \qquad
(n=1,2,3,\ldots),
\end{equation}
subject to the boundary conditions
\begin{equation} \label{zmbcs}
y_n(-1)=0,\quad y_n(1)=0,
\end{equation}
where $R_n(x)$ is defined by
\begin{equation} \label{eq:11}
R_n(x)= {1 \over (n-1)!} {\partial ^{n-1} {\cal N}[\phi(x;p),A(p)]
\over
\partial p^{n-1}} \Big|_{p=0},
\end{equation}
and
\begin{equation*}
\chi_n= \left\{ \begin{array}{ll} 0, & n\leq 1, \\ 1, & n>1.
\end{array} \right.
\end{equation*}
At each iteration, we have two unknowns, $C_1$, in Eq.
(\ref{eq:25}), and $a_n$. These unknowns are obtained by considering
the boundary conditions (\ref{zmbcs}) as follows.

At zero order, we obtain
\begin{equation}
y_0(x)=\tilde y(x).
\end{equation}
At 1th-order, we obtain
\begin{equation}
\label{yuno}
y_1(x)={h\over 2}\left[1-x^2-y_0^2(x)\right]-h\epsilon \int_{-1}^x y_0(t) f(a_0t)dt,
\end{equation}
where we have imposed the condition $y_1(-1)=0$ choosing the
integration constant, $C_1$, appropriately. At this moment $a_0$
is free, but we can fix the value of $a_0$ by imposing $y_1(1)=0$:
\begin{equation}
\label{acero}
\int_{-1}^1 y_0(t) f(a_0t)dt=0.
\end{equation}
This is a non linear equation for $a_0$. When $f(x)$ is an even polynomial of degree $2n$, it is
a polynomial equation of degree $n$ in the variable $a_0^2$. Therefore, at this order, the number
of limit cycles is the number of positive roots of this equation, at most $n$ (see \cite{ll4}
for a longer discussion of this point).

Then, for every one of the $a_0$ solutions of the above equation,
i.e. for every one of the limit cycles of the system, we proceed
order by order in $p$ to obtain higher order corrections to the
amplitude $a$ and to the shape of the limit cycle $y(x)$. At any
$n$th-order, with $n\ge 2$, we obtain the $n$th-order correction
$y_n$ to the shape of the limit cycle:
\begin{equation}
\label{pn}
\begin{array}{ll}
y_n(x)= & \displaystyle{y_{n-1}(x)-{h\over 2}\sum_{k=0}^{n-1}y_k(x)y_{n-k-1}(x)-\epsilon h
\int_{-1}^xy_{n-1}(t)f(a_0t)dt}
\\ &\displaystyle{-\epsilon h\sum_{k=1}^{n-1}\int_{-1}^xy_{n-k-1}(t)\sum_{m=1}^k{B_{k,m}(a_1,a_2,...,a_k)
\over m!}t^mf^{(m)}(a_0t)dt,}
\end{array}
\end{equation}
where $B_{n,m}(a_1,...,a_n)$ are the partial ordinary Bell polynomials [\cite{riordan}, p. 190].
We have $B_{0,0}=1$ and, for $n=1,2,3,...$, $B_{n+1,0}=0$ and they satisfy the recurrence
\begin{equation}
B_{n,m}(a_1,...,a_n)=\sum_{k=m-1}^{n-1}a_{n-k}B_{k,m-1}(a_1,...,a_n).
\end{equation}
We recall that $B_{n,m}(a_1,...,a_n)$ is linear in $a_n$.

At $n$th-order, we can obtain also the $n$th-order correction
$a_{n-1}$ to the amplitude of the limit cycle by imposing the
condition $y_n(1)=0$. The result is a linear equation for
$a_{n-1}$ with solution:
\begin{equation}
\label{an}
\begin{array}{ll}
a_{n-1}= &
-\left[\displaystyle{\int_{-1}^1y_{n-1}(x)f(a_0x)dx+\sum_{k=1}^{n-1}{\sum_{m=1}^k}^*y_{n-k-1}(x)
{B_{k,m}(a_1,...,a_k)\over m!}x^mf^{(m)}(a_0x)dx}\right]\\
& \hskip 5mm\left/\left[\displaystyle{\int_{-1}^1x
y_0(x)f'(a_0x)dx}\right], \right.
\end{array}
\end{equation}
where the star in the sum symbol means that the term $(k,m)=(n-1,1)$ is excluded.

For example, at 2th-order we obtain
\begin{equation}
\label{ydos}
y_2(x)=y_1(x)-hy_0(x)y_1(x)-h\epsilon \int_{-1}^x[y_1(t)f(a_0t)+t y_0(t) f'(a_0t)a_1]dt
\end{equation}
and
\begin{equation}
\label{auno}
a_1=-\left[\int_{-1}^1 y_1(x)f(a_0x)dx\right]\left/\left[\int_{-1}^1 x y_0(x) f'(a_0x)dx\right].\right.
\end{equation}

Therefore, a first approximation to the shape of the limit cycle is $y_0(x)+y_1(x)$,
and a first order approximation to the amplitude of the limit cycle is $a=a_0+a_1$.

At 3th-order, we have
\begin{equation}
\begin{array}{ll}
y_3(x)= & \displaystyle{y_2(x)-hy_0(x)y_2(x)-{h\over 2}y_1^2(x)} \\ & \hskip -1cm\displaystyle{-h\epsilon
\int_{-1}^x\left[y_2(t)f(a_0t)+a_1ty_1(t)f'(a_0t)+{1\over 2}a_1^2t^2y_0(t)f''(a_0t)+t y_0(t) f'(a_0t)a_2\right]dt}
\end{array}
\end{equation}
and
\begin{equation}
\label{ados}
\begin{array}{ll}
a_2= & \displaystyle{-\left[\int_{-1}^1
\left[y_2(x)f(a_0x)+a_1xy_1(x)f'(a_0x)+{1\over
2}a_1^2x^2y_0(x)f''(a_0x) \right]dx\right]} \\ &\hskip 5mm \left/
\displaystyle{\left[\int_{-1}^1 x y_0(x) f'(a_0x) dx\right].}
\right.
\end{array}
\end{equation}
Therefore, a second order approximation to the shape of the limit cycle is $y_0(x)+y_1(x)+y_2(x)$ and
a second order approximation to the amplitude of the limit cycle is
$a=a_0+a_1+a_2$.

Moreover, at every order in $p$ the appropriate function $h(\epsilon)$ is
indicated by the experiment.

\subsection{Results for the van der Pol equation}

For the van der Pol system, we have
$f(x)=x^2-1$, $F(x)=x^3/3-x$, $x_0=-1$, $\tilde a=2$ and
\begin{equation}
\tilde y(x)=\sqrt{1-x^2}+{\epsilon\over 2}\left\lbrace
\begin{array}{ll}
0 \hskip 2cm& {if}\hskip 2mm -1\le x<-1/2, \\
F(2)-F(2x) \hskip 5mm & {if}\hskip 2mm -1/2\le x\le 1.
\end{array}\right.
\end{equation}

The expressions for $y_n(x)$ given by the recursive formula (\ref{pn})
can be calculated with Mathematica. Although they are elementary functions,
their length is excessive to be reproduced here.
From (\ref{acero}) we get
\begin{equation}
\int_{-1}^1 y_0(t) f(a_0t)dt={9\epsilon + 8 \pi\over 64}
(a_0^2-4)=0 \Longrightarrow a_0=2.
\end{equation}

From (\ref{auno}) we obtain
\begin{equation}
a_1(h)=-{27\sqrt{3}\epsilon h\over 35(8\pi+9\epsilon)}.
\end{equation}

Therefore, a first approximation to the amplitude of the limit cycle is
\begin{equation}
\label{vpauno}
a(h)\simeq 2-{27\sqrt{3}\epsilon h\over 35(8\pi+9\epsilon)}.
\end{equation}
At it was advanced at the beginning of this section,
note that this expression is only valid for positive $\epsilon$.
Choosing $b=-1.3$ and $c=4$ in (\ref{ache}),
\begin{equation}
h(\epsilon)=-{1.3\epsilon\over 4+\epsilon^2},
\end{equation}
the formula $a_{R1}(\epsilon)$ given in (\ref{ar1}) is obtained.

From (\ref{ados}) we get
\begin{equation}
  \begin{array}{ll}
& a_2(h)= \\ &\displaystyle{{3 \epsilon h [19197 \sqrt{3} \epsilon^2 h - 5040 (8\sqrt{3}- 21 h) \pi +
2 \epsilon (-22680\sqrt{3} + h (54837 + 7700\sqrt{3}\pi))]\over 19600 (9 \epsilon +
   8 \pi)^2}.}
   \end{array}
\end{equation}

Therefore, a second approximation to the amplitude of the limit cycle is
\begin{equation}
\label{vpados}
  \begin{array}{ll}
a(h)\simeq & \displaystyle{2-{27\sqrt{3}\epsilon h\over 35(8\pi+9\epsilon)}+} \\
& \hskip -15mm\displaystyle{+{3 \epsilon h [19197 \sqrt{3} \epsilon^2 h - 5040 (8\sqrt{3}- 21 h) \pi +
2 \epsilon (-22680\sqrt{3} + h (54837 + 7700\sqrt{3}\pi))]\over 19600 (9 \epsilon + 8 \pi)^2}.}
  \end{array}
\end{equation}
Choosing $b=-0.561$ and $c=4$ in (\ref{ache}),
\begin{equation}
\label{nocheache}
h(\epsilon)=-{0.561\epsilon\over 4+\epsilon^2},
\end{equation}
the formula $a_{R2}(\epsilon)$ shown in (\ref{ar2}) is finally obtained.

As an example, and to conclude this section, we plot in Fig. 3 the form
of the van der Pol limit cycle when $y_0(x)$, $y_0(x)+y_1(x)$ and $y_0(x)+y_1(x)+y_2(x)$
are used as approximated limit cycles. It is not banal to recall here that the reconstruction
of the shape of limit cycles is not a problem less difficult than that of the calculation
of their amplitudes.

\vskip 0.5cm
\begin{center}
\begin{figure}[t]
\centerline{\includegraphics[width=8cm]{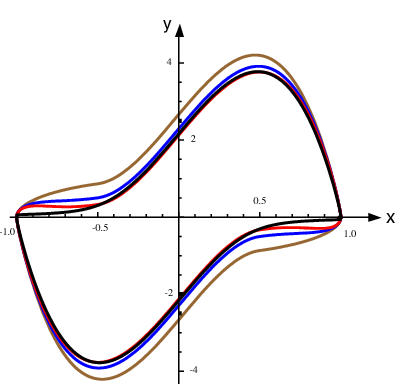}}
\caption{Approximated shape of the van der Pol limit cycle for $\epsilon=5$ and $h(\epsilon)$
given in (\ref{nocheache}). Black line: the experimental limit cycle. Brown, blue and red lines:
the curves $y_0(x)$, $y_0+y_1(x)$ and $y_0+y_1(x)+y_2(x)$, respectively.}
\end{figure}
\end{center}

\section{Conclusions}

In this work, the conjecture (\ref{conjec}) has been posed.
This is a consequence of the different formulas here presented for approximating the amplitude
$a$ of the van der Pol limit cycle. In addition to the well known constant approximation $\bar{a}=2$
that generates an error less than $1\%$, we establish a family of recursive formulas that are valid
for the whole range of the parameter $\epsilon$. Two of them, $a_{R1}(\epsilon)$ and
$a_{R2}(\epsilon)$, have been explicitly given. The first one, $a_{R1}$, produces an error
less than $0.1\%$ and the second one, $a_{R2}$, reduces the error to less than $0.05\%$.
Moreover, $a_{R2}$ is conjectured to be an upper bound of $a$.

As far as we know, this is the first time where an analytical approximation
of the amplitude of the van der Pol limit cycle, with validity from
the weakly up to the strongly nonlinear regime, is proposed.

{\bf Acknowledgements:} 

The {\it Gobierno of Navarra, Spain, Res. 07/05/2008} is acknowledged by its
financial support.

\end{document}